\documentclass[pra,aps,reprint,showpacs,superscriptaddress,amsmath]{revtex4-1}

\usepackage{bm}
\usepackage{graphicx}
\usepackage{dsfont}
\usepackage[utf8]{inputenc}
\usepackage[caption=false]{subfig}

\newcommand{\bra}[1]{\langle #1 | \,}
\newcommand{\ket}[1]{\, | #1 \rangle}

\newcommand{\expv}[1]{\langle #1 \rangle}
\newcommand{\om}{\omega}
\newcommand{\Om}{\Omega}

\newcommand{\ga}{\gamma}

\newcommand{\de}{\delta}
\newcommand{\De}{\Delta}

\newcommand{\br}{\mathbf{r}}
\newcommand{\E}{\hat{\mathcal{E}}}
\newcommand{\bk}{\hat{b}}
\newcommand{\ac}{\hat{a}_c}

\newcommand{\sge}{\sigma_{ge}}

\newcommand{\sse}{\sigma_{se}}
\newcommand{\seg}{\sigma_{eg}}
\newcommand{\ssg}{\sigma_{sg}}
\newcommand{\see}{\sigma_{ee}}
\newcommand{\sss}{\sigma_{ss}}

\begin{document}
\title{A number-state filter for pulses of light}

\author{Nikolai Lauk}
\email{nlauk@physik.uni-kl.de}
\author{Michael Fleischhauer}
\affiliation{Department of  physics and research center OPTIMAS,
University of Kaiserslautern}

\date{\today}

\begin{abstract}
We present a detailed theoretical analysis of a Fock-state filter based on the
photon-number dependent group delay in cavity induced transparency
proposed in Phys. Rev. Lett. \textbf{105}, 013601 (2010). We derive a general
expression for the propagation velocity of different photon-number components
of a light pulse interacting with an optically dense ensemble of
three-level atoms coupled to a resonator mode under conditions
of cavity induced transparency. These predictions are compared to
numerical simulations of the propagation of few photon wave packets, and an estimation for experimental realization is made. 
\end{abstract}


\maketitle

\begin{section}{Introduction}
Creation of non classical states of light is one of the central topics in 
quantum optics. Photon number states, also called Fock states of light are maybe the most prominent 
representatives of those. They are of particular interest for quantum information 
processing as they can be used as a carrier of discrete bits of quantum information
\cite{QInfo,DLCZ,Milburn_2001}. 
Despite the fact that by now there are many successful experimental realizations for creation of single-photon states, e.g. in cavity-QED systems \cite{Haroche_2001,Haroche_2013,Walther_2000,Rempe_2007,Rempe_2009,Martinis_2008,Kuhn_2015}, an ideal single-photon source, a device that efficiently provides propagating single-photons on demand, is still missing.
An extremely useful tool would be a filter that allows to extract different photon-number components of
a propagating wave packet. Such a system was proposed in \cite{Gor_2010}, where a possibility of spatial separation of
different photon-number components of an initially coherent pulse was shown.
However, the theoretical analysis performed in \cite{Gor_2010} was based on nonlinear operator equations. To handle these a couple of simplifying assumptions and approximations on operator level had to be made. These are known to be difficult to justified in general. In the present paper we re-examine this system and provide a rigorous and quantitative analysis of the scheme including an assessment of experimental requirements. 

The proposed Fock-state filter
is based on a phenomenon called cavity induced transparency (CIT). 
It occurs in an ensemble of three level atoms with a $\Lambda$-type
configuration of couplings to two electromagnetic fields and is 
closely related to the well-known effect of electromagnetically 
induced transparency (EIT) \cite{EIT_Rev}. The difference between the two systems is the 
replacement of the coherent control field in EIT by a quantized cavity mode. 
The coupling of the atomic ensemble to the cavity mode, even if it is in the vacuum state, 
can lead to transparency for the propagating probe field in otherwise opaque medium. Transparency induced by an empty cavity, called vacuum induced transparency (VIT),
was theoretically proposed in \cite{VIT_proposal} and has been demonstrated 
experimentally in \cite{vuletic_2011}. The interaction of the probe field with the coupled
atom-cavity system leads to a temporary transfer of photons from the probe field
to the cavity mode. The number of cavity photons is determined by the number of the probe field photons, and therefore is proportional to
the probe field intensity.
The back-action of the hybrid atom-cavity system onto the probe field, in 
particular its effect onto the group velocity, depends on the strength of the cavity field, i.e. on the number of cavity photons.
As a consequence different photon-number
components of the probe field propagate with different velocities causing a photon-number
dependent group delay of the probe field. This process is analyzed in detail in this paper.

The paper is organized as follows. In Sect. II we introduce the model, 
discuss the underlying principle of the system and summarize
the expected results based on intuition. Then in Sect. III we 
derive a general expression for the photon-number dependent group velocity using the concept of the dark states. 
To confirm these results we have performed numerical 
wave-function simulations for up to two photons in the initial
pulse, which we present in Sec. IV. Sec. V discusses consequences for
experimental implementations and Sec. VI gives summary and conclusions.
\end{section}

%%%%%%%%%%%%%%%%%%%%%%%%%%%%%%%%
\begin{section}{Model}
%%%%%%%%%%%%%%%%%%%%%%%%%%%%%%%%

We consider a gas consisting of three level atoms in a $\Lambda$-type configuration (see Fig.\ref{Fig:sketch}b). 
A cavity mode $\ac$ couples the $\ket{s}-\ket{e}$ transition of the atoms, the adjacent 
$\ket{g}-\ket{e}$ transition is coupled to a propagating quantum field described by the slowly varying operator
\begin{align}
\E(\br,t)=\frac{1}{\sqrt{V}}\sum_\mathbf{k}\hat{b}_\mathbf{k}\mathrm{e}^{i(\mathbf{k}\cdot\br-\omega_\mathbf{k}t)}\mathrm{e}^{-i (k_pz-\om_pt)},\label{eq:SVFO}
\end{align} 
where $\hat{b}_\mathbf{k}$ is annihilation operator of a photon in mode $\mathbf{k}$ with corresponding frequency $\om_\mathbf{k}$, $V$ is the quantization volume, and $\om_p$ is the carrier frequency of the quantized probe field with corresponding wave number $k_p=\om_p/c$ (see Fig. \ref{Fig:sketch}a). Initially all atoms are assumed to be in the ground state 
$\ket{g}$ and the cavity mode is in the vacuum state. The Hamiltonian 
of the system in a frame rotating with the probe field carrier frequency $\om_p$ is given by
\begin{align}
\hat{H}&=\sum_{\mathbf{k}}\hbar 
(\om_{\mathbf{k}}-\om_p)\hat{b}_{\mathbf{k}}^\dagger \hat{b}_{\mathbf{k}} + 
\hbar\int d^3\br\  n(\br)\Bigl(\De\see(\br) + \de\sss(\br)\Bigr) \nonumber \\
&-\int d^3\br\ n(\br)\Bigl(g\sge(\br)\E^\dagger (\br) + G \sse(\br)\ac^\dagger + 
H.c.\Bigr),
\label{eq:Hamiltonian}
\end{align}
where we introduced continuous atomic operators $\sigma_{lm}(\br)=\frac{1}{\De N}\sum_j^{\De N} \ket{l_j}\bra{m_j}$ by averaging over small volume around $\br$ containing $\De N$ particles \cite{DSP_2002}. $n(\br)$ is the atomic density, and $\De$ and $\de$ denote the single photon and two photon detuning respectively. $g=d_{ge}\sqrt{\frac{\om_p}{2\hbar\epsilon_0}}$ and $G=d_{se}\sqrt{\frac{\om_c}{2\hbar\epsilon_0V_c}}$ are the single atom coupling constants for the probe field and cavity field respectively, where $d_{ge}$ and $d_{se}$ are dipole matrix elements of the $\ket{g}-\ket{e}$ and $\ket{s}-\ket{e}$ transitions, and $V_c$ is the cavity quantization volume.

The first line in \eqref{eq:Hamiltonian} describes the free evolution of the atomic system and 
the propagating  probe field and the second line describes the interaction between them.
Note that the free Hamiltonian of the cavity mode vanishes in the chosen rotating frame.

%%%%%%%%%%%%%%%%%%%%%%%%%%%%%%%%%
\begin{figure}[htb]
\includegraphics[width=0.45\textwidth]{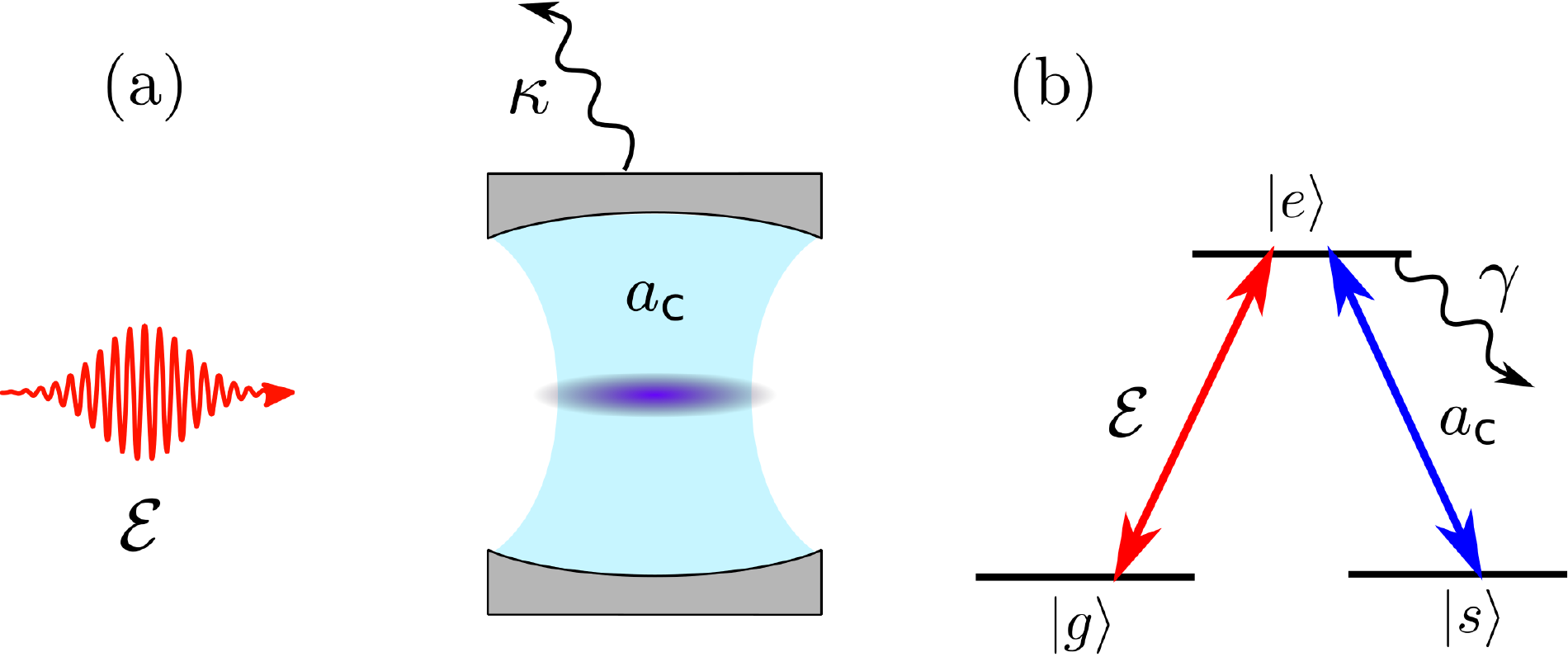}
\caption{(Color online) Set-up of cavity-induced transparency (a) with corresponding atomic level scheme (b).
The probe field ${\E}$ and cavity mode $\hat a_c$ couple to the transitions $|g\rangle-|e\rangle$ and
$|s\rangle -|e\rangle$ from the atomic ground state $|g\rangle$ and meta-stable "spin" state $|s\rangle$
to a common excited states $|e\rangle$ in two-photon resonance.
}\label{Fig:sketch}
\end{figure}
%%%%%%%%%%%%%%%%%%%%%%%%%%%%%%%%%

To get some intuition for the operation of the Fock-state filter let us first consider the EIT case. There 
the atomic transition $\ket{s}-\ket{e}$ is coupled by the classical driving 
field with Rabi frequency $\Omega$. This coupling induces transparency for the 
probe field on the otherwise opaque transition $\ket{g}-\ket{e}$. In addition, 
the group velocity of the probe field is modified \cite{DSP_2002} according to
\begin{align}
\frac{v_g}{c} = \frac{\Omega^2}{g^2n+\Omega^2},
\end{align}
and depends on the strength of the control field $\Om$ and on the collective atom-field coupling $g\sqrt{n}$. 
Now by replacing the driving field with a cavity it seems natural that the 
group velocity will depend on the effective atom-cavity coupling 
$G\sqrt{N}$ where $N$ is the number of photons in the cavity.
Thus we expect that for strong back-action of the atom-cavity system onto the probe field, which happens in the strong coupling regime, i.e. for a single-atom cooperativity 
$C=G^2/\gamma\kappa > 1$, different photon number component will propagate with 
different group velocities. Here $\kappa$ and $\gamma$ are the decay rates of the cavity and the atomic polarization, respectively.
\end{section}

%%%%%%%%%%%%%%%%%%%%%%%%%%
\begin{section}{group velocity}
%%%%%%%%%%%%%%%%%%%%%%%%%%%

To become more acquainted with the system and to introduce the key concept of dark states let us first consider a related toy model, where we consider the probe field as a single mode cavity field $\E = \hat b/\sqrt{V}$. For the sake of simplicity we set all detunings to zero.
The corresponding Hamiltonian is then given by
\begin{align}
 \hat{H}=-\hbar\sum_{i=1}^{N_a} (g\sge^i\frac{\hat b^\dagger}{\sqrt{V}} + G \sse^i\hat{a}^\dagger_c + H.c.)\label{Eq:single}
\end{align}
where the sum runs over all interacting atoms $N_a$.
%
%%%%%%%%%%%%%%%%%%%%%%%%%%%%%%%%%%%
\begin{figure}[h]
\includegraphics[width=0.45\textwidth]{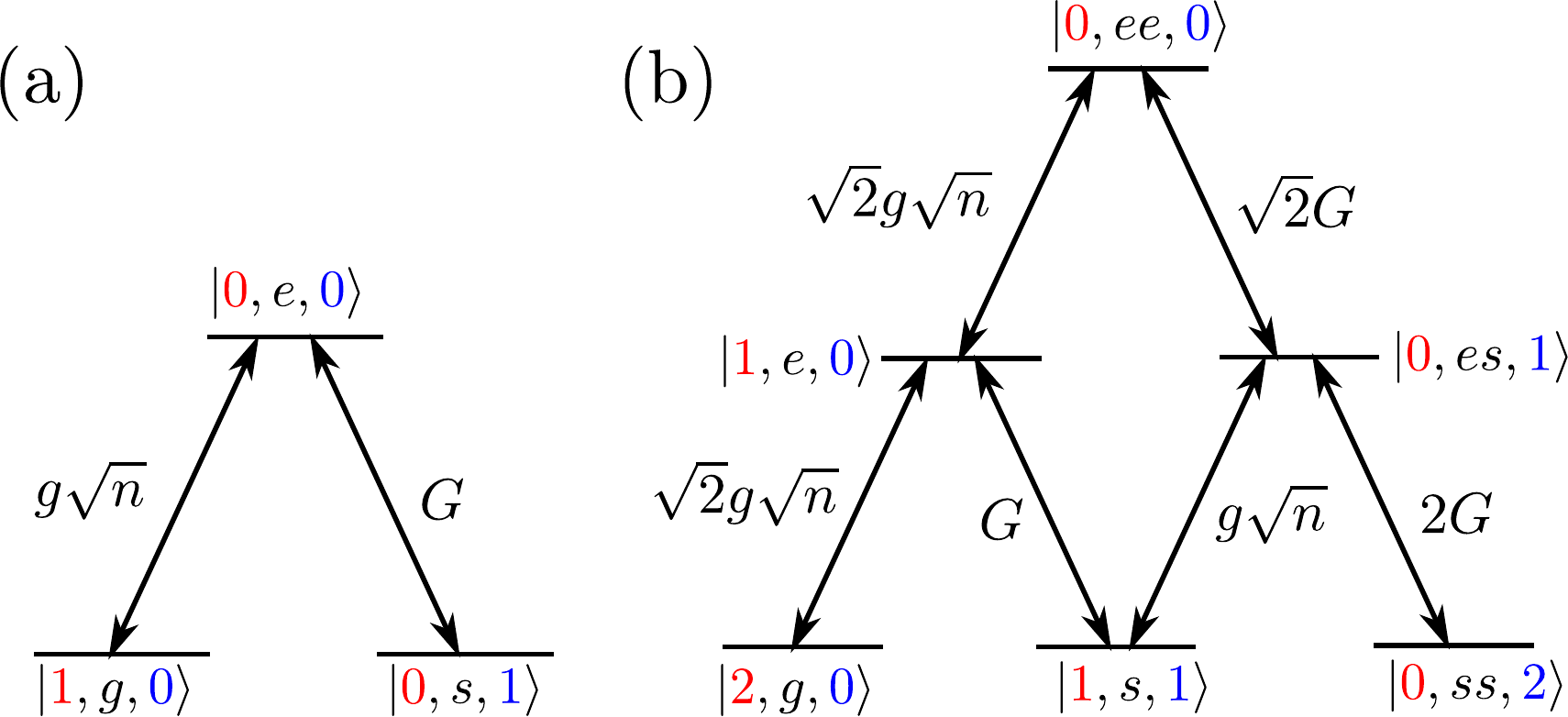}
\caption{(Color online) Participating levels and corresponding coupling schemes for the single (a) and two excitation (b) manifolds.}\label{Fig:coupling}
\end{figure}
%%%%%%%%%%%%%%%%%%%%%%%%%%%%%%%%%%%
%
It is easy to verify that this Hamiltonian conserves the total number of 
excitations, i.e. the Hilbert space splits into decoupled manifolds each of which contains all states with fixed excitation number
(Fig. \ref{Fig:coupling}), and we can treat each manifold separately.
By looking on the spectrum of \eqref{Eq:single} in different manifolds we note that all of these manifolds have in common an existence of an eigenstate with eigenvalue zero, so called dark state.
It is convenient to introduce the following notation for the interacting states 
$\ket{n_s,e^ls^m,n_c}$, where $n_s$ and $n_c$ denote the number of the probe field 
photons and cavity photons respectively and $e^ls^m$ denotes the atomic state 
that interacts with the probe and cavity fields and contains $l$ atoms in 
state $\ket{e}$, $m$ atoms in state $\ket{s}$ and all the other atoms in the 
ground state $\ket{g}$.
Using this notation we can express the dark state in the 
single excitation manifold as
\begin{align}
\ket{\psi_D^{(1)}}=\frac{G}{\sqrt{G^2+g^2n}}\ket{1_s,g,0_c}-\frac{g\sqrt{n}}{
\sqrt{G^2+g^2n}}\ket{0_s,s,1_c}.
\end{align}
Here $g\sqrt{n}$ is the collective coupling strength of $N_a$ atoms in a volume $V$ with homogeneous density $n=N_a/V$ to the mode $\hat{b}$. The dark state for the subspace containing two excitations is given by
\begin{align}
\ket{\psi_D^{(2)}}=\frac{1}{\mathcal{N}}(&\sqrt{2}G^2\ket{2_s,g,0_c}-2G 
g\sqrt{n}\ket{1_s,s,1_c}\nonumber \\
&+g^2n\ket{0_s,ss,2_c})\label{Eq:2dark}
\end{align}
where $\mathcal{N}=\sqrt{2G^4+4G^2g^2n+g^4n^2}$ is the normalization constant. 

Note that in the low excitation limit, i.e. if the number of excitations is much smaller than the number of atoms, atomic excitation can be treated as a bosonic excitation. As a consequence, the coupling of the state $\ket{0_s,ss,2_c}$ to the upper state  $\ket{0_s,es,1_c}$ experiences a \textit{two-fold} bosonic enhancement leading to a term  $2G$ instead of  $\sqrt{2}G$. 
Because of this \emph{two-fold} enhancement we can not write 
the double-excitation dark state as a direct product of two single-excitation dark states.
This is different from usual EIT, where the quantization of the control field is not 
considered and dark states can be represented as number states of a
polariton operator \cite{DSP_2002}.

The general expression for the dark state in the $N$ excitations subspace reads
\begin{align}
\ket{\psi_D^{(N)}}=\frac{1}{\mathcal{N}}\sum_{M=0}^N f^Ms^{N-M} 
\ket{M_s,s^{N-M},(N-M)_c}\label{Eq:Ndark},
\end{align}
where $\mathcal{N}$ is normalization constant and the coefficients are given by
\begin{align}
f^Ms^{N-M} = 
(-1)^M\frac{N!}{(N-M)!}\frac{1}{\sqrt{M!}}\left(\frac{G}{g\sqrt{n}}\right)^M.
\end{align} 
Taking into account a possible decay from the excited atomic state $\ket{e}$ we find another feature of the dark states. Due to fact that all of them do not contain contributions from the atoms in the excited state $\ket{e}$, the dark states are not affected by the decay from this state, which is the origin of their name. As a consequence, the dark states make up stationary states of the system in that case.

Let us proceed and consider the propagation of the probe field. To describe the propagation we
have to include many modes with different wave numbers $\mathbf{k}$, i.e. for a probe field propagating in the $z$-direction we can write  
$\E(z)=\frac{1}{\sqrt{V}}\sum_k \bk_k\mathrm{e}^{ikz}$. Replacing the single mode operator in \eqref{Eq:single} by this expression leads to a modification of the Hamiltonian according to
\begin{align}
\hat{H}=&-\hbar\sum_{i=1} g\sge^i\frac{1}{\sqrt{V}}\sum_k\bk_k^\dagger \mathrm{e}^{-ikz} + G 
\sse^i\hat{a}_c^\dagger + H.c.\nonumber \\
	&+\sum_k \hbar\omega_k \bk_k^\dagger\bk_k\label{Eq:many}
\end{align}
where the additional part corresponds to the energy of the free probe field and 
gives rise to field propagation. We assume here an infinitely extended medium and ignore boundary effects.
In this case the system is translationally invariant and the Hamiltonian (\ref{Eq:many}) does not couple modes
with different $k$'s. 

We start again with a single excitation. Due to translational invariance we can treat all $k$-modes independently,
i.e. for every mode $k$ we have three states $\ket{1_k,g,0_c}$, 
$\ket{0_k,e_k,0_c}$ and $\ket{0_k,s_k,1_c}$ coupled in a $\Lambda$ configuration 
(compare Fig. \ref{Fig:coupling}), where we modified the previous notation by 
labeling it with the mode wave number $k$. Similar to the single mode case 
we can write for the dark state of mode $k$
\begin{align}
\ket{\psi_D^k}=\frac{G}{\sqrt{G^2+g^2n}}\ket{1_k,g,0_c}-\frac{g\sqrt{n}}{\sqrt{
G^2+g^2n}}\ket{0_k,s_k,1_c}.
\end{align}
States belonging to different $k$'s are linearly independent and thus the dark 
state for the entire single-excitation manifold is given by 
\begin{align}
 \ket{\psi_D}=\sum_k C_k\ket{\psi_D^k} \label{Eq:SingleDS}
\end{align}
with some constants $C_k$ which fulfill the normalization condition $\sum_k |C_k|^2 =1$.

To calculate the group velocity we assume that the spectral width of the incoming photon is smaller than the transparency window  $c\De k_s=\De\om_s < \om_{tr}$, which is defined by $\om_{tr}=\frac{G^2}{\ga\sqrt{\mathrm{OD}}}$ \cite{DSP_2002}. Here $\mathrm{OD}=L/L_{\rm abs}$ is the optical depth and $L_{\mathrm{abs}} = \gamma c/g^2n$ is the resonant absorption length of the medium. Since within the transparency width each $k$ excitation is described by the corresponding dark state $\ket{\psi_D^k}$, fulfilling this condition ensures that the entire excitation will propagate as a dark state. In this case we can treat the propagation perturbatively 
by calculating the first order energy correction resulting from the last term in (\ref{Eq:many}).
The group velocity is then given by
\begin{align}
 v_g^{(1)}=\frac{\partial 
\bra{\psi_D^k}\sum_{k'}\omega_{k'}\bk^\dagger_{k'}\bk_{k'}\ket{\psi_D^k}}{\partial 
k}=c\frac{G^2}{G^2+g^2n},
\end{align}
where we used the fact that for a free field $\omega_k=ck$.
%
%%%%%%%%%%%%%%%%%%%%%%%%%%%%%%%%%%
\begin{figure}[h]
 \includegraphics[width=0.4\textwidth]{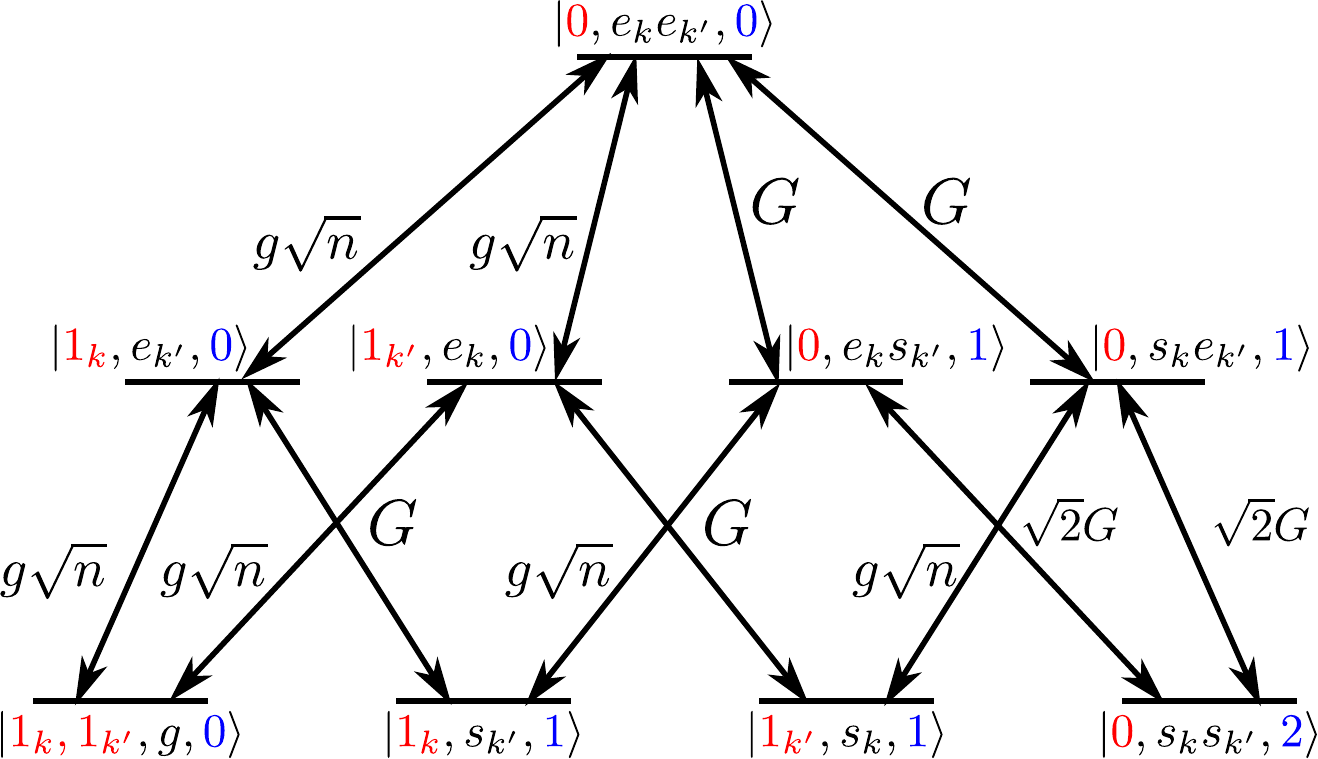}
 \caption{(Color online) Bare levels and coupling diagram in the case of two propagating excitations.}\label{Fig:2coupling}
\end{figure}
%%%%%%%%%%%%%%%%%%%%%%%%%%%%%%%%%%

In the next step we consider two excitations. The corresponding 
coupling scheme is shown in Fig. \ref{Fig:2coupling}. At first glance it looks 
different to the corresponding coupling diagram in the single mode case (see 
Fig. \ref{Fig:coupling}). For example, the state that contains two photonic 
excitations $\ket{1_k,1_{k'},g,0}$ couples now to two different states which 
contain an atomic excitation, namely $\ket{1_k,e_{k'},0}$ and 
$\ket{1_{k'},e_k,0}$. Also the coupling constants are changed. 
However, if we combine these two states to a symmetric superposition state 
$\ket{1,e,0}_S^{kk'}=1/\sqrt{2}(\ket{1_k,e_{k'},0}+\ket{1_{k'},e_k,0})$ and 
apply this procedure to all other denegerate states, we again end up with a
coupling scheme that is identical to that of 
the single mode case. Most importantly the resulting effective coupling constants
between the symmetric states are the same as in the single-mode case.
The corresponding dark state is then given by equation 
(\ref{Eq:2dark}) and reads
\begin{align}
\ket{\psi_D^{kk'}}&=\frac{1}{\mathcal{N}}\Bigl(\sqrt{2}G^2\ket{1_k1_{k'},g,0_c}
+g^2n\ket{0_s,s_ks_{k'},2_c}\nonumber \\
&-2G g\sqrt{n}\frac{1}{\sqrt{2}}(\ket{1_k,s_{k'},1_c}+\ket{1_{k'},s_{k},1_c})\Bigr).
\end{align}
A general dark state containing two excitations then reads
\begin{align}
 \ket{\Psi_D^{(2)}}=\sum_{kk'}C_{kk'}\ket{\psi_D^{kk'}},
\end{align}
in complete analogy to equation (\ref{Eq:SingleDS}). The group velocity can be
determined in a similar manner as in the single-excitation case 
\begin{align}
 v_g^{(2)}&=\frac{\partial \bra{\psi_D^{kk'}}\sum_{\tilde{k}}\omega_{\tilde 
k}\bk^\dagger_{\tilde k}\bk_{\tilde k}\ket{\psi_D^{kk'}}}{\partial K}\nonumber\\
 &=c\frac{2G^4+2G^2g^2n}{g^4n^2+4G^2g^2n+2G^4},
\end{align}
where the differentiation is now made with respect to the center of mass 
momentum $K=k+k'$. 

The generalization to $N$ excitations is straight forward. Start with 
the $N$ excitation dark state (\ref{Eq:Ndark}), replace the state 
$\ket{M_s,s^{N-M},(N-M)_c}$ by the symmetric state of all $\binom{N}{M}$ 
degenerate states. Calculate the first order energy correction and differentiate 
this with respect to the center of mass momentum $K=k_1+k_2+\dots+k_N$ to obtain
the group velocity. 
This yields
\begin{align}
\frac{v_g^{(N)}}{c}=\frac{\sum\limits_{M=0}^N 
\frac{M}{N}\left(f^Ms^{N-M}\right)^2}{\sum\limits_{M=0}^N 
\left(f^Ms^{N-M}\right)^2}.
\label{eq:v-full}
\end{align}
The factor $\frac{M}{N}$ in the nominator results from the  symmetrization procedure 
and can be interpreted as a weighting factor of the corresponding state 
to the group velocity. This means for example that the component of the state containing $N$ photons contributes fully and 
the symmetric state with only one photon contributes with relative weight $1/N$ to 
the propagation velocity.

The dependence of the group velocity on the number of incoming photons according to equation (\ref{eq:v-full}) is plotted in Fig. 
\ref{Fig:groupvelocity}. One notices that the group velocity for $N$ photons 
is always smaller than the group velocity for $N+1$ photons and that in the 
limit of large $N$ the group velocity approaches the vacuum speed of light, as one
would expect.

%%%%%%%%%%%%%%%%%%%%%%%%%%%%%%%%%%%%%%%%
\begin{figure}[h]
\includegraphics[width=0.45\textwidth]{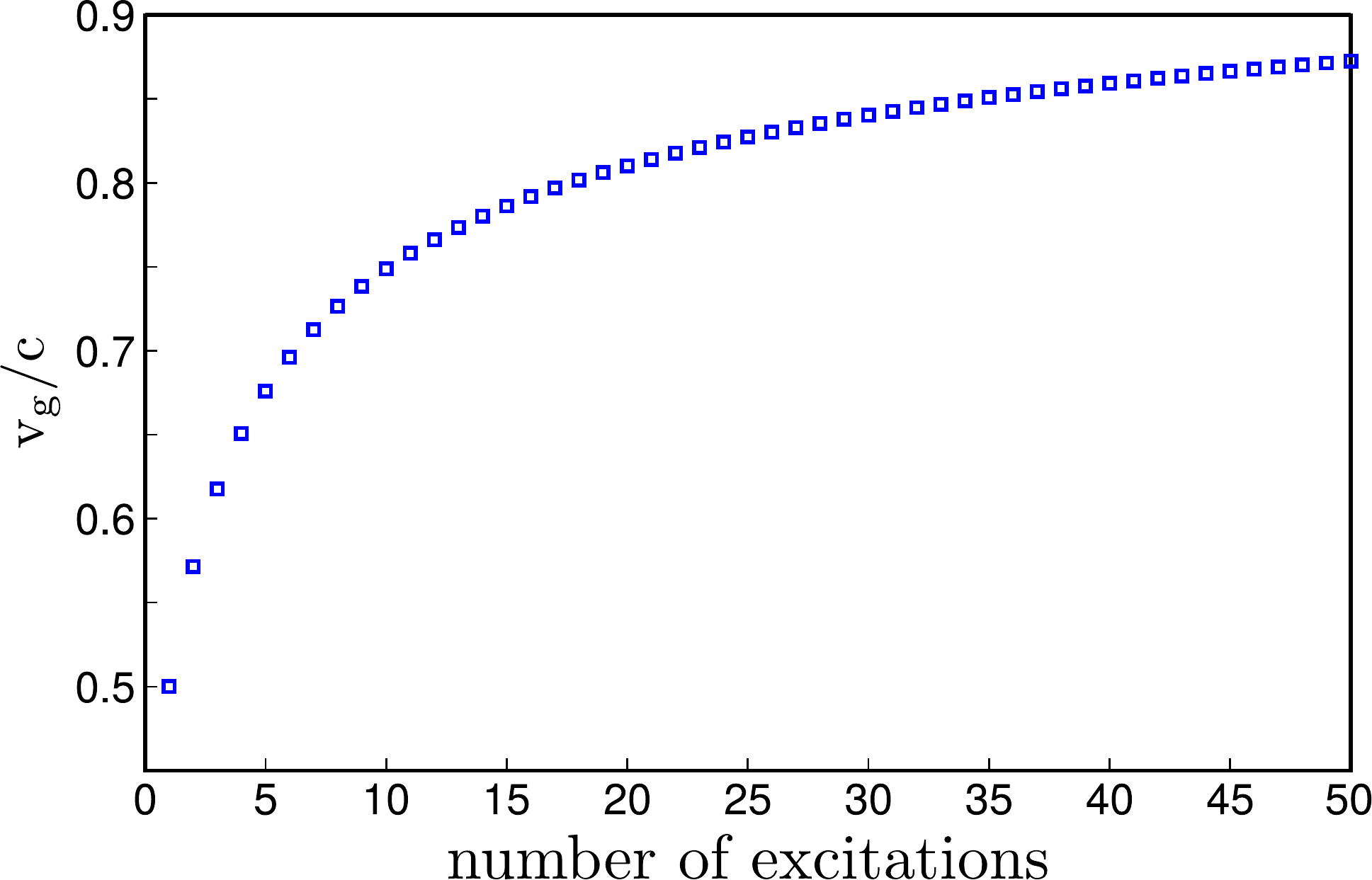}
\caption{(Color online) The dependence of the group velocity on the number of incoming photons in the case $G$=$g\sqrt{n}$. The group velocity for some fixed photon number $N$ is always larger than the one for the preceding numbers and approaches the vacuum speed of light in the limit of large photon numbers $N$.}\label{Fig:groupvelocity}
\end{figure}  
%%%%%%%%%%%%%%%%%%%%%%%%%%%%%%%%%%%%%%%%%

%%%%%%%%%%%%%%%%%%%%%%%%%%%%%%%%%%%%%%%%
\begin{figure}
\includegraphics[width=0.45\textwidth]{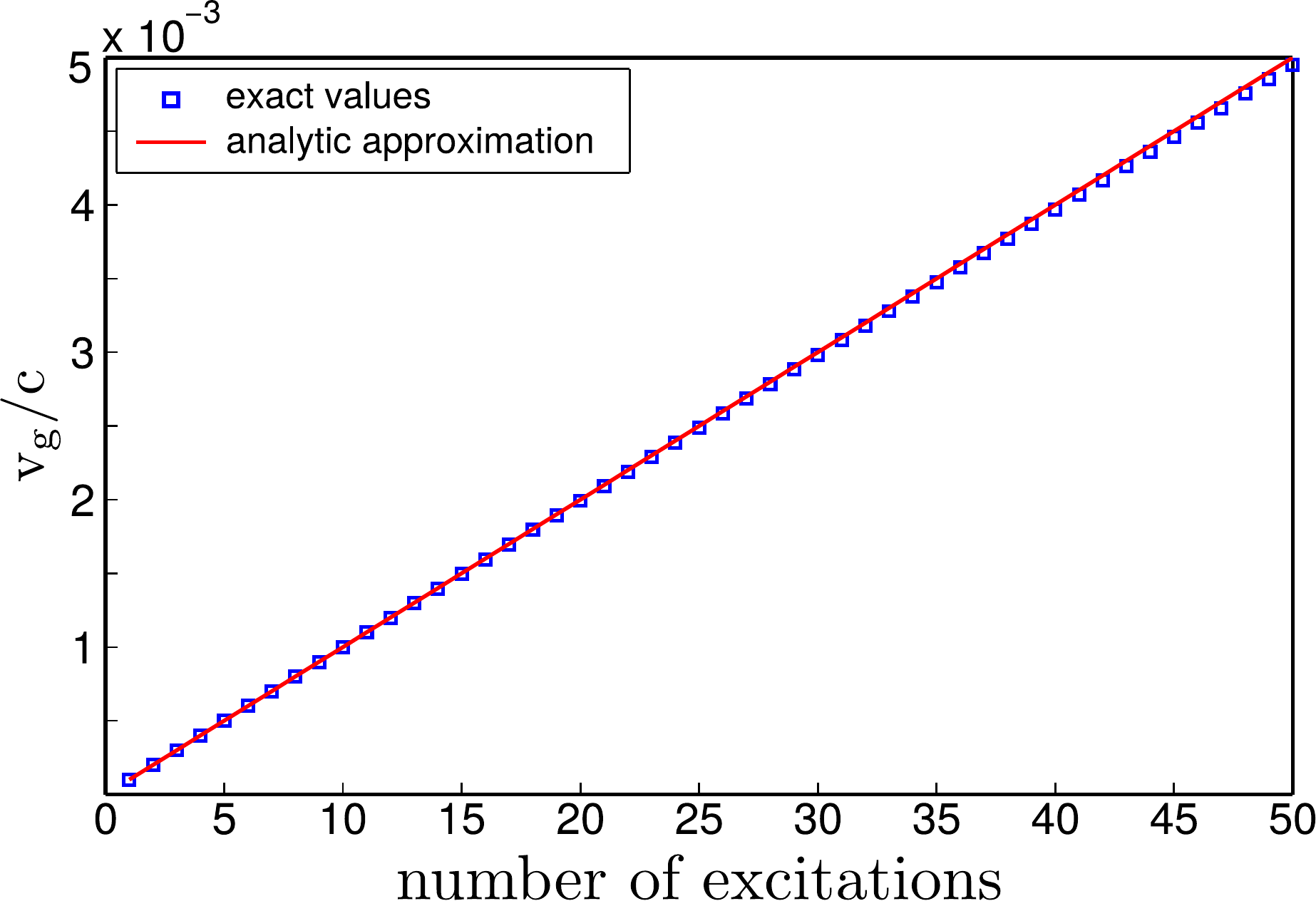}
\caption{(Color online) The dependence of the group velocity on the number of incoming photons in the case $G$=$0.01g\sqrt{n}$ and the corresponding fit of the approximation (\ref{Eq:approx}).}\label{Fig:approx}
\end{figure}  
%%%%%%%%%%%%%%%%%%%%%%%%%%%%%%%%%%%%%%%%%

In the limit $G\ll g\sqrt{n}$, which is typically the case in
experiments, we can give an analytic approximation for the group velocity
\begin{align}
\frac{v_g}{c} \approx \frac{G^2}{g^2n}N,\label{Eq:approx}
\end{align}  
i.e. the group velocity scales linearly with the number of photons.
A comparison between this approximation and the full expression
(\ref{eq:v-full}) is shown in Fig. \ref{Fig:approx}.
\end{section}

%%%%%%%%%%%%%%%%%%%%%%%%%%%%%%%%%%%%%%%
\begin{section}{Numerical results}
%%%%%%%%%%%%%%%%%%%%%%%%%%%%%%%%%%%%%%%

To confirm the results derived in the previous section and to take into account
boundary effects associated with finite spatial extend of the medium we numerically simulate
the propagation of pulses with up to two photons.
We perform the simulations using Hamiltonian (\ref{eq:Hamiltonian}) by making a wave function ansatz and numerically integrating
the corresponding Schr\"odinger equations for the amplitudes of the different components. 
The single excitation wave function reads
\begin{align}
\ket{\psi(t)}=&\int \! d^3\br \frac{f(z,t)}{\sqrt{V}}\E^\dagger(\br)\ket{0}+\int 
\! d^3\br\sqrt{n(\br)}\frac{e(z,t)}{\sqrt{V}}\sigma_{eg}(\br)\ket{0}\nonumber \\
&+\int 
d^3\br\sqrt{n(\br)}\frac{s(z,t)}{\sqrt{V}}\sigma_{sg}(\br)a^\dagger_c\ket{0},
\end{align}
where $|f(z,t)|^2$ corresponds to probability of finding a photon at
position $z$, and the probability of finding an atom at the same position in state $\ket{e}$ and 
$\ket{s}$ is given by $|e(z,t)|^2$ and $|s(z,t)|^2$ 
respectively. Using the commutator relations
\begin{align}
 [\E(\br),\E^\dagger(\br')]&=\delta(\br-\br'),
\end{align}
\begin{align}
 [\sigma_{ij}(\br),\sigma_{kl}(\br')]&=\frac{1}{n(\br)}(\delta_{jk}\sigma_{il}(\br)-\delta_{il}\sigma_{kj}(\br))\delta(\br-\br')
\end{align}
we obtain the corresponding equations of motion
\begin{align}
 \partial_t f(z,t) &= -c\partial_z f(z,t) +ig\sqrt{n(\br)}e(z,t),\\
 \partial_t e(z,t) &= -\ga e(z,t)\!+i g\sqrt{n(\br)}f(z,t)+i G^*\! s(z,t),\\
 \partial_t s(z,t) &= i G e(z,t),
\end{align}
where we include the decay from the excited state $\ket{e}$. These equations are equivalent to the propagation equations in the EIT case. From the EIT case we know that if the pulses are long enough, i.e. if they fulfill the adiabaticity condition $T_p>1/\om_{tr}$ \cite{DSP_2002}, we can adiabatically eliminate the $e(z,t)$ component. This allows us to recast the equations of motion to a single propagation equation for the $f(z,t)$ component, namely 
\begin{align}
(\partial_t + v_g\partial_z) f(z,t)=0 
\end{align}
i.e. the single photon pulse travels through the medium without being absorbed with reduced group 
velocity $v_g=cG^2/(G^2+g^2n)$, which coincides with the propagation velocity $v_g^{(1)}$ derived
in the last section for the single excitation. The corresponding time delay after propagation reads
then $\De\tau=L/v_g^{(1)}$, where $L$ is the medium length. Using the EIT analogy we can also describe the behavior of the pulse
on the medium boundary, where the group velocity changes from $c$ to $v_g$. Such a
change leads to a pulse compression inside the medium by the factor $v_g/c$.

Let us move on to two-photon pulses. 
%%%%%%%%%%%%%%%%%%%%%%%%%%%%%%%%%%%%%%%%%%%%%%%%%%
\begin{figure}[htp]
\centering
\subfloat[Initial Gaussian wave packet $ff(z_1,z_2)=\frac{1}{\sqrt{\pi}}\mathrm{e}^{-2(z_1-2)^2-2(z_2-2)^2}$ ]{
\includegraphics[width=\columnwidth]{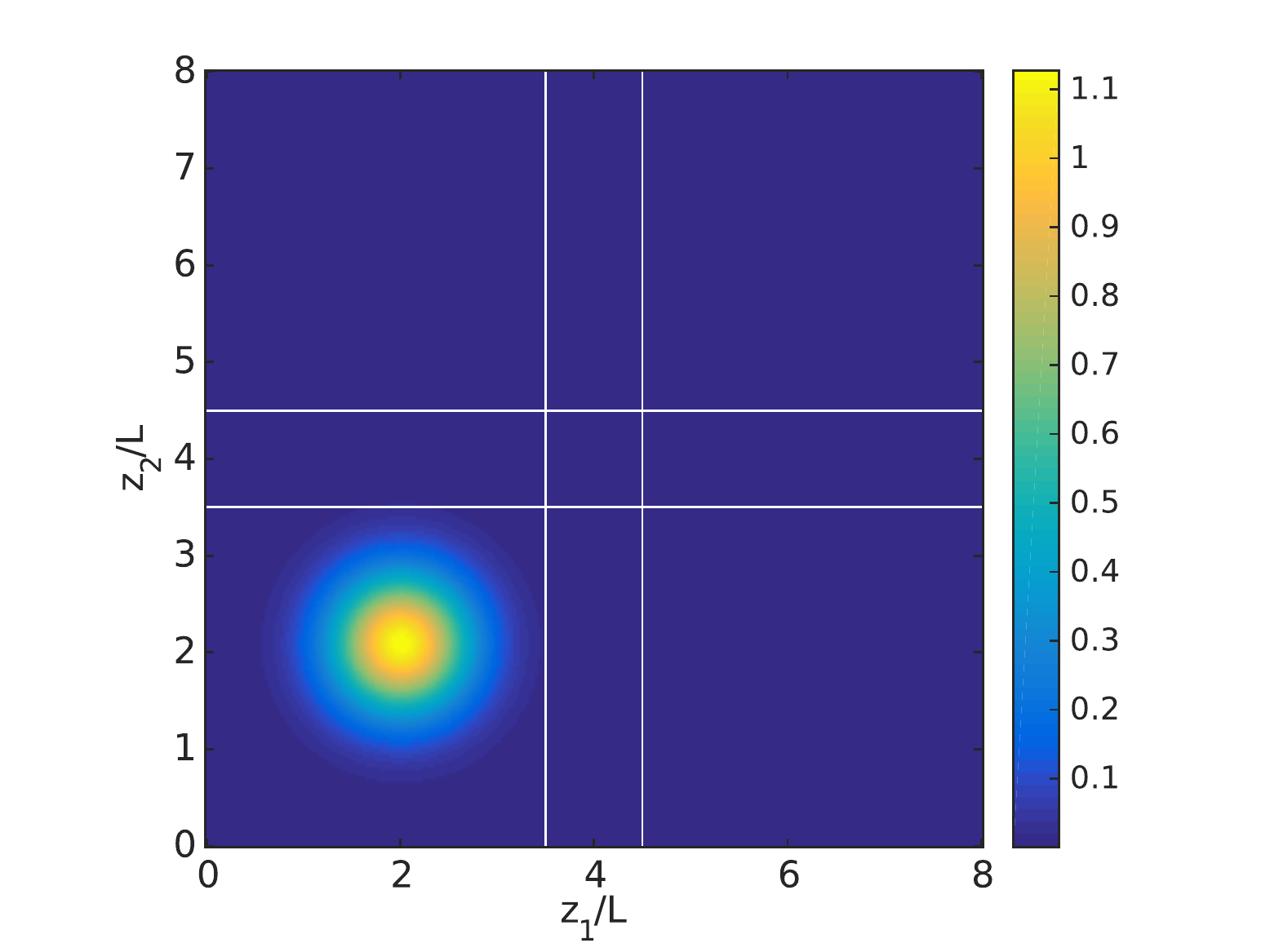}}

\subfloat[The two photon component after propagating inside the medium. One notices the spatial compression of the pulse. ]{
\includegraphics[width=\columnwidth]{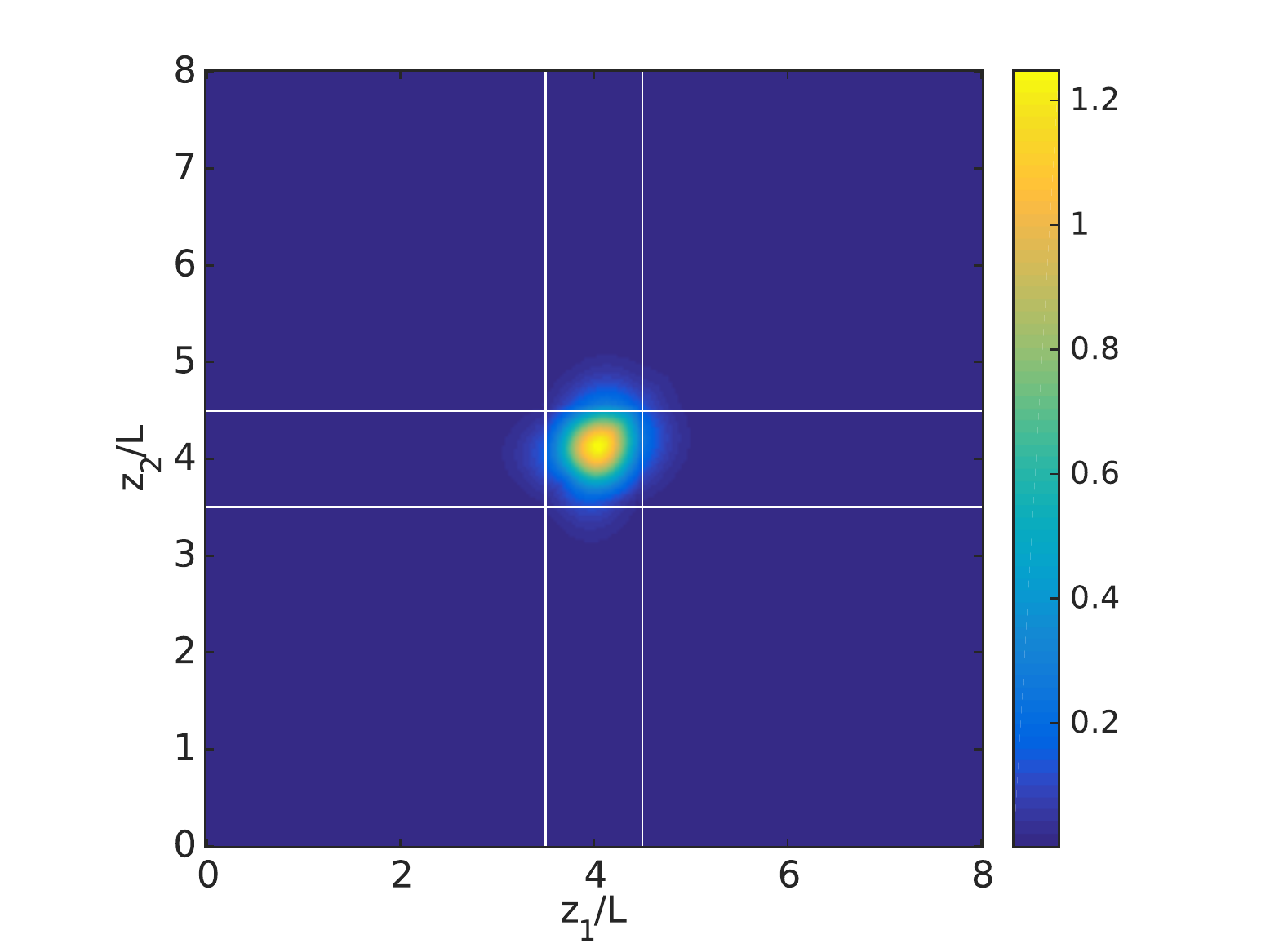}}

\subfloat[After the propagation through the medium one recognizes the distortion of the pulse shape due to non vanishing mutual distance between the two photons.]{
\includegraphics[width=\columnwidth]{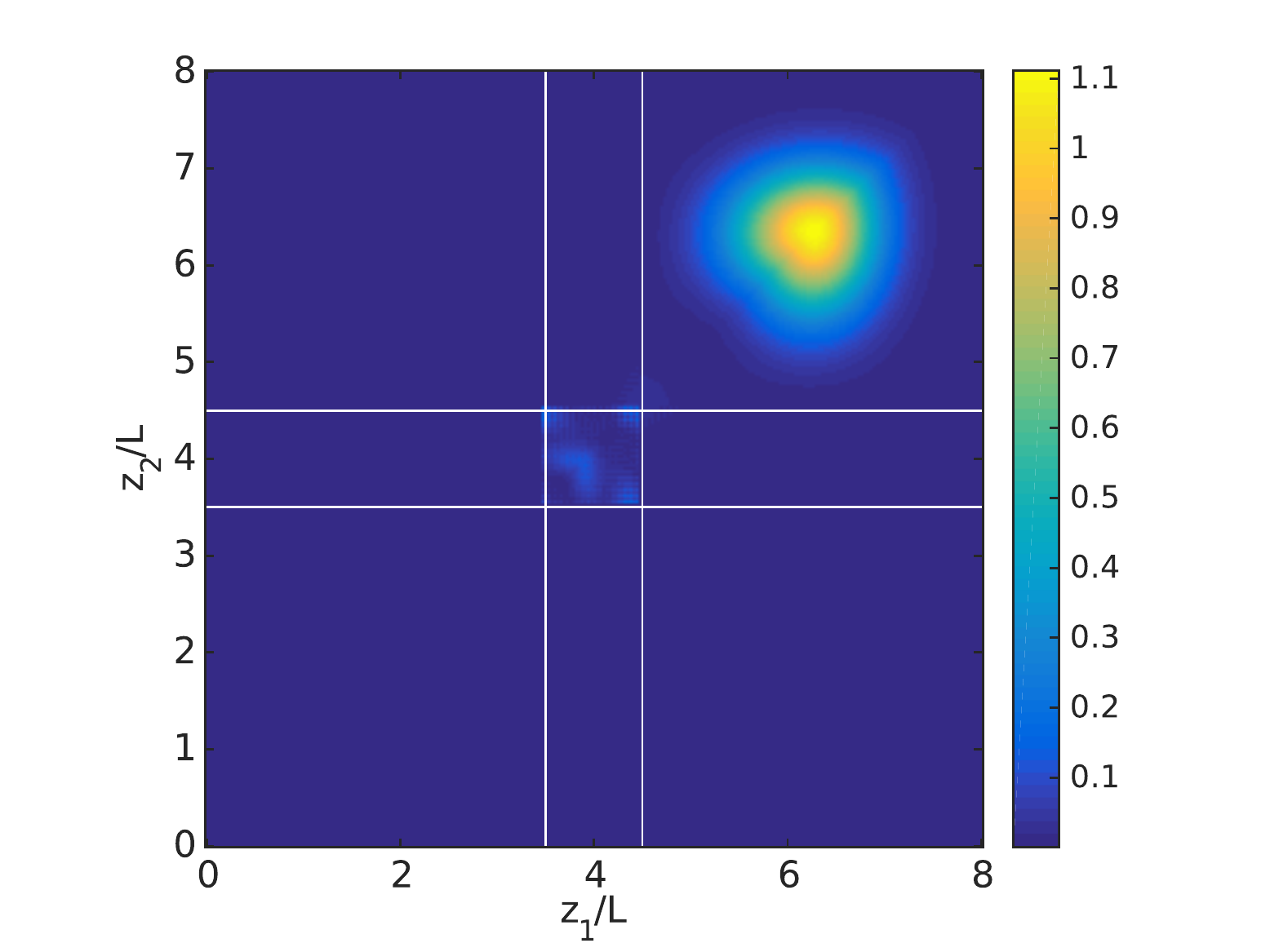}}
\caption{The time evoluton of the two photon component $ff(z_1,z_2)$ of an initial Gaussian pulse. The parameters are $G=g\sqrt{n}=500 c/L$. The white lines denote the medium boundaries.}\label{Fig:twophoton}
\end{figure}
%%%%%%%%%%%%%%%%%%%%%%%%%%%%%%%%%%%%%%%%%%%%%%%%%%%%%
Here the wave function can be written as
\begin{align}
&\ket{\psi(t)}=\frac{1}{\sqrt{2}}\int d^3\br \int d^3\br^\prime 
\frac{ff(z,z',t)}{V}\E^\dagger(\br)\E^\dagger(\br^\prime)\ket{0}\\ \nonumber
					  &+\int d^3\br \int d^3\br^\prime 
\frac{ef(z,z',t)}{V}\sqrt{n(\br)}\seg(\br)\E^\dagger(\br')\ket{0}\\ \nonumber
					  &+\frac{1}{\sqrt{2}}\int d^3\br \int 
d^3\br^\prime 
\frac{ee(z,z',t)}{V}\sqrt{n(\br)}\sqrt{n(\br')}\seg(\br)\seg(\br')\ket{0}\\ 
\nonumber
					  &+\int d^3\br \int d^3\br^\prime 
\frac{sf(z,z',t)}{V}\sqrt{n(\br)}\ssg(\br)a^\dagger_c\E^\dagger(\br')\ket{0}\\ 
\nonumber
					  &+\int d^3\br \int d^3\br^\prime 
\frac{es(z,z',t)}{V}\sqrt{n(\br)}\sqrt{n(\br')}
\seg(\br)\ssg(\br')a^\dagger_c\ket{0}\\ \nonumber
					  &+\frac{1}{2}\int d^3\br \int 
d^3\br^\prime 
\frac{ss(z,z',t)}{V}\sqrt{n(\br)}\sqrt{n(\br')}\ssg(\br)\ssg(\br'){a^\dagger_c}
^2\ket{0} .\nonumber
\end{align}
Similar to single excitation the absolute value squared of the coefficients gives the probability of finding 
the system in the corresponding state.
In Fig. \ref{Fig:twophoton} we plot the quantity $|ff(z,z')|^2$, which is proportional to the probability of 
finding two photons at positions $z$ and $z'$. We see that just as in the single-photon
case the two-photon pulse is compressed inside the medium. However, in addition
we recognize that the shape of the wave function is distorted after propagation
through the medium. To understand how this distortion comes 
about let's consider some component $ff(z_1,z_2)$. As already mentioned the 
absolute value squared gives the probability of finding two photons with mutual 
distance $d=|z_1-z_2|$. Initially both photons are outside of the 
medium and travel with the speed of light. Then the first photon enters the medium and 
propagates now with the reduced group velocity $v_g^{(1)}$ until after the time 
$t=d/c$ the second photon enters the medium. Since now there are two photons 
inside the medium they both propagate with the group velocity $v_g^{(2)}>v_g^{(1)}$.
Due to pulse compression in the medium the distance of the two photons is reduced to $d'=dv_g^{(1)}/c$.
Then after the time 
$t'=(L-d')/v_g^{(2)}$, where $L$ is the medium length, the first photon will 
leave the medium and the remaining photon will now propagate with the group 
velocity $v_g^{(1)}$ until it leaves the medium. Afterwards both photons 
will again propagate with the speed of light. This shows that the amount of 
time that both photons propagate with the two-photon group velocity $v_g^{(2)}$ depends on 
their mutual distance inside the medium $d'$. Taking this into account we can 
explain the shape distortion. The components on the first bisectrix have the smallest 
possible distance $d=0$ and travel at all times with the larger group velocity 
$v_g^{(2)}$ and hence are more advanced in comparison to other components with 
non vanishing mutual distance. The maximal time delay between the single and two 
photon component is then
\begin{align}
\De\tau_{12}=L\left(\frac{1}{v_g^{(1)}}- \frac{1}{v_g^{(2)}}\right)\approx 
\frac{1}{2}L\frac{g^2n}{G^2}=\frac{\ga}{2G^2}\mathrm{OD},
\end{align}
where $\mathrm{OD}$ is the optical depth of the medium.
The other extreme case is when the mutual distance between two photons inside the medium becomes larger than 
the medium size $L$. Obviously these components propagate only with the velocity
$v_g^{(1)}$ and therefore can not be separated from the single photon components. 
This puts a limitation on the maximal pulse length. On the other hand one can not use 
arbitrary short pulses, since those would violate the adiabaticity condition and 
lead to pulse absorption. Rewriting the adiabaticity condition in terms of 
maximal delay time we can give an upper bound for the ratio of the maximal delay 
time to the pulse time
\begin{align}
 \frac{\De\tau_{12}}{T_p}<\frac{1}{2}\sqrt{\mathrm{OD}}.
\end{align}
In order to be able to effectively separate the single photon component this ratio should be larger than 1.
Both conditions can only be satisfied at large optical depths.

At the end of this section we want to make some remarks on the pulses containing more than two photons. Since the dimension of the Hilbert space grows exponentially with the number of excitations it is clear that the wave function ansatz becomes unattractive for more than two photons. However, we can use the mutual distance argument also in the case of multiple excitations by taking into account all possible distances between photons, e.g. the three photon component $fff(z_1,z_2,z_3)$ will propagate with the group velocity $v_g^{(3)}$, iff the largest mutual distance is smaller than the medium, i.e. all three photons are inside the medium. The group velocity will be $v_g^{(2)}$ if only two photons are present in medium either due to the transition from free space to the medium or because the largest mutual distance is larger than the medium. For all other cases the component will propagate with the velocity $v_g^{(1)}$. In principle this procedure can be generalize for $N$ photon component resulting in a complicated bookkeeping for the all possible distances. However, if one is mainly interested in the separation of the single photon component from the rest it is enough to consider single and two photon components, since as we see from Fig. \ref{Fig:groupvelocity} the group velocity for higher components is also higher. That means that if one manages to resolve the single photon component from the two photon component it will be automatically separated from the other components, too.
\end{section}
\begin{section}{Estimation for experimental realization}
In this section we want to investigate the possibility for an experimental 
realization of our proposal. State of the art cavities can reach single atom 
coupling strength of about $G\approx(2\pi)3 {\rm MHz}$ with cavity decay rates of roughly 
$\kappa\approx(2\pi)0.1 {\rm MHz}$ \cite{Donner_2014}. Using a Bose-Einstein condensate as our three level 
medium allows us to obtain the required optical depths. For example using Rb BEC 
one can reach optical depths of $\mathrm{OD}\approx(10 - 100)$ \cite{Esslinger_2007,highOD} and a single atom 
cooperativity of $C\gtrsim15$ \cite{Donner_2014,Reichel_2007}. A weak laser pulse can be used as the propagating probe field,
i.e. we can approximate the state of the incoming field as
\begin{align}
 \ket{\alpha}&\approx 
\left(1-\frac{|\alpha|^2}{2}\right)\ket{0}+\alpha\int\!dzf(z)\E^\dagger(z)\ket{0
}+\nonumber \\
&\frac{\alpha^2}{2}
\int\!dz_1\int\!dz_2f(z_1)f(z_2)\E^\dagger(z_1)\E^\dagger(z_2)\ket{0},
\end{align}
which is a good approximation for a weak coherent pulse.
\begin{figure}[h]
\centering
\includegraphics[width=0.45\textwidth]{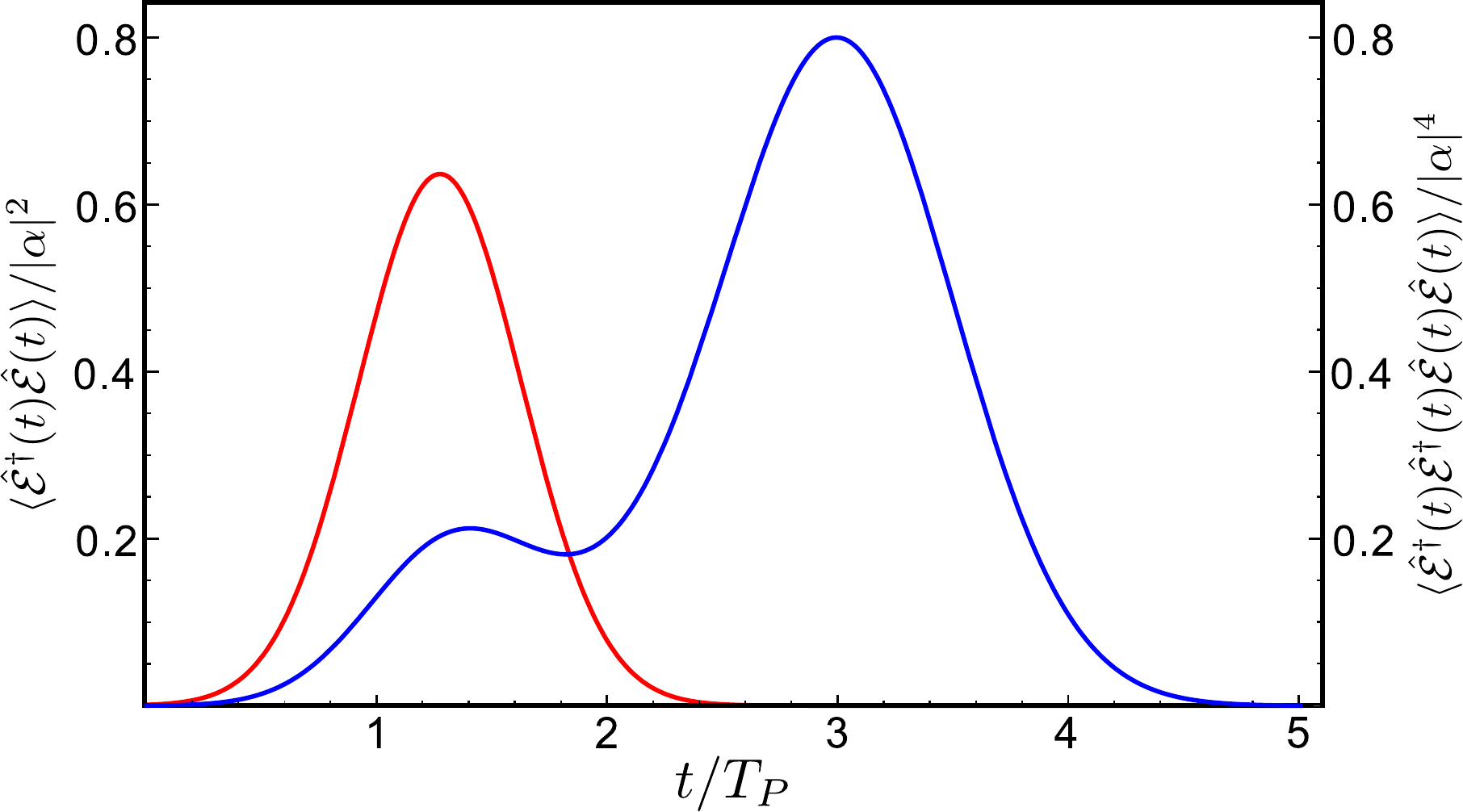}
\caption{(Color online) Intensity (blue) and two photon component (red) of the field after 
propagation. The calculation are performed for a weak coherent pulse $T_p=1\mathrm{\mu s}$ containing on averange $\bar{n}=|\alpha|^2=0.25$ photons. The other parameters are $G=(2\pi)3 {\rm MHz}$ and optical depth $\mathrm{OD}=50$.}\label{Fig:experiment}
\end{figure}
This allows us to utilize the results of our calculations and extract all 
relevant quantities. In Fig. \ref{Fig:experiment} we plot the intensity 
$\expv{\E^\dagger(t)\E(t)}$ of the field after propagation through the medium calculated using experimental realistic numbers from above and setting all detunings to zero.
Already in this intensity plot we can recognize the spatial separation of different components. To make this separation more evident we also 
plot the expectation value of the two photon component $\expv{\E^\dagger(t)\E^\dagger(t)\E(t)\E(t)}$,
here we can clearly see that this component is about $\De\tau_{12}$ ahead of the single photon component. 

Until now, we completely disregarded cavity damping and the 
excited state decay in our considerations. While we can safely neglect the excited state decay as 
long as we fulfil the adiabaticity condition, the cavity decay could be a 
practical limitation, since it will destroy the dark states and induce coupling 
between different excitation manifolds. Its influence can be neglected if the 
cavity lifetime is larger than the propagation time of the single photon, since 
it is the slowest component, i.e.
\begin{align}
 \kappa<\frac{v_g^{(1)}}{L}\approx\frac{G^2}{\gamma \mathrm{OD}}.
\end{align}
This condition can be rewritten in terms of the cavity cooperativity leading 
to more restrictive condition $C>\mathrm{OD}$ on the cavity than the usual strong coupling 
condition $C>1$. This represents a major limitation for the experimental realization.
\end{section}

%%%%%%%%%%%%%%%%%%%%%%%%%%%%%%%%%%%%%%%
\begin{section}{conclusion}
%%%%%%%%%%%%%%%%%%%%%%%%%%%%%%%%%%%%%%%

In conclusion, we presented a detailed analysis of our proposal for a number state filter for propagating light pulses based on cavity induced transparency. Assuming adiabaticity and an infinite homogeneous medium we derived a general expression for the dependence of the group velocity on the number of incoming photons. To take into account the effects associated with the finite medium size we performed numerical simulations for few-photon wave packets. Using the results of these simulations we could explain the behavior of the light pulse components with different photon numbers at the medium boundaries and derive a condition for the separation of the single photon component from the rest. Finally we investigated a possibility for an experimental realization of our proposal. We found that for successful implementation we have to modify the usual strong coupling condition in terms of the cavity cooperativity $C>1$ to the more restrictive condition $C>\mathrm{OD}$, where $\mathrm{OD}$ is the optical depth of the medium.
\end{section}

%%%%%%%%%%%%%%%%%%%%%%%%%%%%%%%%%%%%%%%
\begin{acknowledgments}
%%%%%%%%%%%%%%%%%%%%%%%%%%%%%%%%%%%%%%%
The authors would like to thank Razmik Unanyan for fruitful discussions.
\end{acknowledgments}


%\bibliographystyle{apsrev4-1}
%\bibliography{VIT-paper}

\end{document}